\documentclass[12point]{article}
\usepackage{hyperref}
\usepackage{amsmath}
\usepackage{graphicx}
\usepackage{amssymb}

\usepackage{hyperref}

\renewcommand{\title}[1]{%
    \bigskip%
    \begin{center}%
    \Large\bf #1%
    \end{center}%
    \vskip .2in}

\renewcommand{\author}[1]{%
    {\begin{center}
    #1
    \end{center}}}
\newcommand{\address}[1]{\vspace{-1.7em}\vspace{0pt}
    {\begin{center}
    \it #1
    \end{center}}}

\begin{document}


\title{Constrained Hamiltonian analysis of a non relativistic Schrodinger field coupled with C-S gravity}

\author
{
,
{Pradip Mukherjee $\,^{\rm a,b}$} and  Abdus Sattar  $\,^{\rm c,d}$}

\address{$^{\rm a}$Department of Physics, Barasat Government College,\\10, KNC Road, Barasat, Kolkata - 700124.\vspace*{1cm}

\address{$^{\rm c}$Tapna Chaturbhuj High School\\ Balakhali Tapna, Bishnupur, South 24 pgs, West Bengal-743503, India }\vspace{0.5em}

 }\vspace{0.5em}

\address{$^{\rm b}$\tt mukhpradip@gmail.com}
\address{$^{\rm d}$\tt sat.abdus@gmail.com}

\begin{abstract}
We provide a  constrained Hamiltonian analysis of a non relativistic Schrodinger field in 2+1 dimensions , coupled with Chern - Simons gravity. The coupling is achieved by the recently advanced Galilean gauge theory \cite{BMM1},\cite{ BMM2}, \cite{BM4}. The calculations are repeated with a truncated model to show that deviation from  Galilean gauge theory makes the theory untenable. The issue of nonrelativistic spatial diffeomorphism is discussed in this context.
\end{abstract}

\section{Introduction} 
Diffeomorphism of the spacetime manifold is in itself not a physical symmetry; the physics is determined by the spacetime symmetry in the locally inertial manifold \cite{W}. In this sense we talk of relativistic or nonrelativistic diffeomorphism invariance. 
Non relativistic diffeomorphism invariance  (NRDI) has recently gained considerablely interest in the literature \cite{SW,Bekaert:2011qd,Hoyos:2011ez, Schaefer:2013oba,Hoyos:2013qna} due to its diverse application in condensed matter physics (specifically in the theory of fractional quantum hall effect)(FQHE),hollographic models \cite{Janiszewski:2016zrm}, Newtonian Gravity and others. It was none other than Cartan  \cite{Cartan-1923,Cartan-1924} who formulated a geometric theory of Newtonian gravity way back in 1923 . Much work was done \cite{Havas, ANDE, EHL, MALA, MTW} on the geometric properties of the corresponding Newton - Cartan (NC) spaceime. However. during resurgence of the NRDI the chief issue was coupling of non relativistic field theories with background curved spacetime \cite{SW}, which was not much discussed in the then literature. A host of applications of the NRDI model of \cite{SW} appeared in the literature \cite{Bekaert:2011qd,Hoyos:2011ez, Schaefer:2013oba,Hoyos:2011ez}. However. certain problems appeared in the formulation of \cite{SW}. These are,
\begin{enumerate}
\item
The transformation of the metric becomes non canonical and

\item
Galilean symmetry could not be retrieved in the flat limit

\end{enumerate}
The problems were tackled by considering a gauge field and relating the Galilean boost parameter with the gauge parameter. Assuming a $U(1)$ gauge field in the context of FQHE is only natural. But trading off galilean boost symmetry with $U(1)$ gauge symmetry is not very apetizing. Again, that this endeavour decreases the number
 of symmetry elements was overlooked. Following this line of research, a U(1)
gauge field was later introduced as an element of NC geometry \cite{J}. The geometric structure erected by a long work of many stalwarts in the field  was thus required to be modified. 
Different
 approaches to the problem, namely the algebraic method \cite{abpr}, coset construction \cite{Karananas:2016hrm}, nonrelativistic limit procedures \cite{J} and others evolved to investigate NRDI but it can be asserted that a general procedure for coupling nonrelativistic field theories with gravity was not available.
  
 In this scenario Galilean gauge theory (GGT) \cite{BMM1,BMM2,BMM3,BM4} was formulated basing on the gauging of symmetry approach introduced by Utiyama \cite{U} for relativistic theories, tailored appropriately for nonrelativistic theories.
  Spatial diffeomorphism can be easily obtained from GGT \cite{BM4}. However there are significant differences in some issues between the result from GGT with other approaches.This is most prominant in the coupling of the Schrodinger field theory with curve space (\cite{SW}) where Galilean symmetry can only be retrieved in the flat limit if there is a gauge field (see above). On the other hand the 
  spatially diffeomorphic theory obtained from GGT finds smoothly the flat galilean limit and  does not require any additional gauge interaction. Following  the GGT approach one can consistently tackle the issue of torsion in Newton Cartan space time 
 \cite {BM5} or provide  the basis for Milne boost symmetry of metric NC theory \cite{BM6}, to name a few examples, within the purview of the NC geometry. However, the dynamical consistency of GGT is yet to be examined.
 Naturally, Hamiltonian analysis is an important tool to understand the consistency of a field theoretic model. The objective of this work
 is to formulate the Schrodinger field coupled with gravity as obtained from GGT
in the phase space. Note that there are very few examples of such analysis available in the literature, still fewer with the motivation of the present work.

  Hamiltonian structure of non relativistic Schrodinger model coupled with curved space time as obtained from GGT will be analysed here. Observe that so far  we consider
theories  coupled with background gravity. Interestingly, symmetries of a  model with background interaction which are evident from the action  can not be reproduced by Hamiltonian method. 
 For the latter, dynamics of the gravitational interaction  is required to be included. This is not surprising because Hamiltonian analysis is performed in the phase space where the variables are coordinates and their conjugate momenta.The latter is derived by differentiating the Lagrangian with respect to generalised velocity. The momenta conjugate to the background fields weakly vanish. In the Hamiltonian framework these are constraints. Conservation of these constraints is the step where dynamics comes into play. However, when fields do not have any dynamics, such analysis is bound to be trivial.
 

Consequently, for useful Hamiltonian analysis, we will have to supplement the action obtained from GGT with a dynamical term for gravity. Now in 2+1 dimension the Chern Simons term provides an interesting dynamical term for both relativistic and non relativistic models. Thus Chern - Simons gravity \cite{Witten} will be a suitable choice. The fields appearing in our model have origin in the localisation process. It thus necessarily contains Hamiltonian constraints. A comprehensive method of Hamiltonian analysis for such singular system was introduced by Dirac \cite{D}. Our aim is to analyse Chern Simons gravity coupled  non relativistic schrodinger field model by Dirac's method and to discuss the consistency of the model. This will enable us to compare different spatially diffeomorphic models also, as we will see. We will provide a comprehensive account of constraints structure of the model in question which is a novel calculation. The inclusion of the Chern Simons gravity in the context of spatial diffeomorphism is once again a unique feature. No doubt that the problem investigated in this paper is quite interesting in its own merit. 

Before finishing the introductory section an account of the organisation of the paper will be appropriate. In the next section the nonrelativistic Schrodinger field theory coupled with background gravity is written from GGT.  As we have learnt, the dynamics of gravity must be included in  our model to carry out a  meaningful Hamiltonian analysis. 
In $(2+1)$ dimensions the  Chern Simons gravity action is a simple and very important candidate for the dynamics. The Chern Simons
gravity action is introduced and its reduction in the adapted coordinates is discussed. Adding the piece with the first part from GGT the complete action is obtained. The Hamiltonian analysis is presented in section 3. This Hamiltonian analysis is repeated in the next section with a truncated action which manifests a magical change of the results. We see that it leads to unphysical degree of freedom of the system. In the next section the results are discussed in the context of
 the present state of the art. Section 6 contains the concluding remarks.

\section{The model}
The Galilean gauge theory (GGT) enables us to couple a nonrelativistic field theory with background gravity \cite{BMM1}, \cite{BMM2}.
The free Schrodinger field theory in galilean coordinates is given by
\begin{eqnarray}
S = \int d^3x\left[ \frac{i}{2}\left(\psi^*\partial_0\psi - \psi\partial_0\psi^*  \right) -\frac{1}{2m}\partial_k\psi^* \partial_k\psi\right]\label{fs}
\end{eqnarray}
where $\psi$ and $\psi^
*$ are the complex Schrodinger fields.

According to GGT, to derive the corresponding coupled action we have to replace the partial derivatives $\partial_\mu\psi$ by the corresponding
$\nabla_\mu\psi$ 
where
\begin{eqnarray}
\nabla_0\psi &=& \Sigma_0{}^\sigma \left(\partial_\sigma + i B_\sigma \right)\psi\nonumber\\
\nabla_a\psi &=&\Sigma a{}^l \left(\partial_l + i B_l \right)\psi\label{kd}
\end{eqnarray}
 $\Sigma$ and $B$ fields, originally introduced as compensating (gauge) fields, are identified with the vierbein and spin connection of the Newton Cartan spacetime \cite{BMM1,BMM2}. If  $\sigma_{ab}, mx_a $ are the generators of spatial rotation and Galileo boost.
\begin{equation} 
B_{\mu}=\frac{1}{2}B^{ab}_{\mu}\sigma_{ab}+B^{a0}_{\mu}mx_{a}
\label{bstructure}
\end{equation}
The last equation introduces the independent fields $B^{a0}_{\mu}$ and $B^{ab}_{\mu}$ which, along with 
$\Sigma_\alpha{}^\mu$ constitute the configuration space of the theory. Note that there is an asymmetry in the expression of the covariant derivative, $\Sigma_a{}^0 =0 $
but  $\Sigma_0{}^k\ne 0 $. Also $B_\mu ^{0a} =0 $ while $B_\mu ^{a0} \ne 0 $
. These are reflection of the fact that time and space are treated in different ways in nonrelativistic 
physics.

From (\ref{fs}), following the procedure detailed above and correcting for the measure we get the action of Schrodinger field coupled with  background Newtonian gravity.  The Lagrangian density becomes \cite{BMM1, BM4},
\begin{eqnarray}
S = \int d^3x \det{\Sigma_\alpha{}^\mu}\left[ \frac{i}{2}\left(\psi^*\nabla_0\psi - \psi\nabla_0\psi^*  \right) -\frac{1}{2m}\nabla_a\psi^* \nabla_a\psi\right]\label{slcompact}
\end{eqnarray}
Expanding, we get
\begin{multline}
\label{ssg}
\mathcal{L}=\frac{M}{\Sigma^{0}_{0}}\Bigl[\frac{i}{2}\Sigma^{0}_{0}\left(\psi^{*}\partial_{0}\psi-\psi\partial_{0}\psi^{*}\right)+\frac{i}{2}\Sigma^{k}_{0}\left(\psi^{*}\partial_{k}\psi-\psi\partial_{k}\psi^{*}\right)\\
-\Sigma^{0}_{0}B_{0}\psi^{*}\psi-\Sigma^{k}_{0}B_{k}\psi^{*}\psi-\frac{1}{2m}\Sigma^{k}_{a}\Sigma^{l}_{a}\left(\partial_{k}\psi^{*}-iB_{k}\psi^{*}\right)\left(\partial_{l}\psi+iB_{l}\psi\right)\Bigr]
\end{multline}

An important point may be emphasised about the Hamiltonian analysis of (\ref{ssg}). 
In this theory $ \Sigma $ and $B $ are background fields,introduced originally as compensating gauge fields and later identified as the vielbeins and spin connections respectively . From the Hamiltonian point of view these fields act like Lagrange multipliers and not as dynamical fields. They are thus not included in the phase space variables. As a result the symmetries exhibited by the action do not show up in the Hamiltonian analysis. Meaningful Hamiltonian analysis is possible when an appropriate kinetic term is provided to define the dynamics.
 We chose  2+1 dimensional Chern-Simons term to make the fields dynamical. The Chern Simons term being a topological term, does not have an independent dynamics. Thus it may be coupled both with relativistic and non relativistic theories. Also the Churn Simons gravity is a very important part in $(2+1) -$ dim gravity. So, the  Hamiltonian analysis presented here has genuine intrinsic appeal.
   
  The 
Lagrangian for the  Chern-Simons gravity is
\begin{equation}
{\mathcal{L}_{cs}}=\epsilon^{\gamma\lambda\rho}\Lambda^{\alpha}_{\gamma}R_{\alpha\lambda\rho}
\end{equation}
where
 \begin{eqnarray}
 R_{\alpha\lambda\rho}=\partial_{\lambda}\omega_{\alpha\rho}-\partial_{\rho}\omega_{\alpha\lambda}+\epsilon_{\alpha\beta\gamma}\omega^{\beta}_{\Lambda}\omega^{\gamma}_{\rho}\label{R} 
 \end{eqnarray} 
and
 \begin{eqnarray}
 \omega_{\alpha\rho}=-\frac{1}{2}\epsilon_{\alpha\beta\gamma} B^{\beta\gamma}_{\rho}\label{W}
 \end{eqnarray}
 In order to write the appropriate action in the Galilean frame in Newton Cartan spacetime, we have to substitute $\Sigma_a{}^0 = 0$ and $ B_\mu{}^{0a}= 0 $ \cite{BMN}.
 
From (\ref{R}) and (\ref{W}), we have
\begin{align*}
R_{0\lambda\rho}&=-\frac{1}{2}\epsilon_{ab}\left(\partial_{\lambda}B^{ab}_{\rho}-\partial_{\rho}B^{ab}_{\lambda}+\frac{1}{2}B^{a0}_{\rho}B^{b0}_{\lambda}\right)\\
R_{a\lambda\rho}&=-\frac{1}{2}\epsilon_{ab}\partial_{\lambda}B^{b0}_{\rho}+\frac{1}{2}\epsilon_{ab}\partial_{\rho}B^{b0}_{\lambda}-\frac{1}{4}\epsilon_{cd}\left(B^{a0}_{\lambda}B^{cd}_{\rho}-B^{a0}_{\rho}B^{cd}_{\lambda}\right)\label{•}
\end{align*}

Using the expressions of $R_{0kl}$, $R_{akl}$ and $R_{a0l}$ we can write the C-S piece as,
\begin{multline*}
\mathcal{L}_{cs}=-\frac{1}{2}\epsilon^{kl}\epsilon_{ab}\Lambda^{0}_{0}\left(\partial_{k}B^{ab}_{l}-\partial_{l}B^{ab}_{k}+\frac{1}{2}B^{a0}_{k}B^{b0}_{l}\right)\\
+\epsilon^{kl}\Lambda^{a}_{0}\Bigl[-\frac{1}{2}\epsilon_{ab}\partial_{k}B^{b0}_{l}+\frac{1}{2}\epsilon_{ab}\partial_{l}B^{b0}_{k}-\frac{1}{4}\epsilon_{cd}\left(B^{a0}_{k}B^{cd}_{l}-B^{a0}_{l}B^{cd}_{k}\right)\Bigr]\\
-2\epsilon^{kl}\Lambda^{a}_{k}\Bigl[-\frac{1}{2}\epsilon_{ab}\partial_{0}B^{b0}_{l}+\frac{1}{2}\epsilon_{ab}\partial_{l}B^{b0}_{0}-\frac{1}{4}\epsilon_{cd}\left(B^{a0}_{0}B^{cd}_{l}-B^{a0}_{l}B^{cd}_{0}\right)\Bigr]
\end{multline*}

After adding Chern-Simons gravity term, the dynamically complete  Lagrangian density is given by
\begin{equation}
\mathcal{L} = \mathcal{L}+\mathcal{L}_{cs}
\end{equation}. 
Explicitly, in terms of the basic fields $\psi$, $\psi^*$, $\Sigma$ and $B$,we have,
\begin{multline}
\label{wl}
\mathcal{L} = \frac{M}{\Sigma^{0}_{0}}\Bigl[\frac{i}{2}\Sigma^{0}_{0}\left(\psi^{*}\partial_{0}\psi-\psi\partial_{0}\psi^{*}\right)+\frac{i}{2}\Sigma^{k}_{0}\left(\psi^{*}\partial_{k}\psi-\psi\partial_{k}\psi^{*}\right)\\
-\Sigma^{\mu}_{0}B^{a0}_{\mu}mx_{a}\psi^{*}\psi-\frac{1}{2m}\Sigma^{k}_{a}\Sigma^{l}_{a}\left(\partial_{k}\psi^{*}-iB^{b0}_{k}mx_{b}\psi^{*}\right)\left(\partial_{l}\psi+iB^{c0}_{l}mx_{c}\psi\right)\Bigr]\\
-\epsilon^{kl}\Lambda^{0}_{0}\frac{\epsilon_{ab}}{2}\left(\partial_{k}B^{ab}_{l}-\partial_{l}B^{ab}_{k}+\frac{1}{2}B^{a0}_{k}B^{b0}_{l}\right)+\epsilon^{kl}\Lambda^{a}_{0}\Bigl[\frac{\epsilon_{ab}}{2}\left(\partial_{l}B^{b0}_{k}-\partial_{k}B^{b0}_{l}\right)-\frac{\epsilon_{cd}}{4}\left(B^{a0}_{k}B^{cd}_{l}-B^{a0}_{l}B^{cd}_{k}\right)\Bigr]\\
-2\epsilon^{kl}\Lambda^{a}_{k}\Bigl[\frac{\epsilon_{ab}}{2}\left(\partial_{l}B^{b0}_{0}-\partial_{0}B^{b0}_{l}\right)-\frac{\epsilon_{cd}}{4}\left(B^{a0}_{0}B^{cd}_{l}-B^{a0}_{l}B^{cd}_{0}\right)\Bigr]
\end{multline}
where
we propose to analyse the constraint structure of the theory (\ref{wl}), using Dirac's method of constrained Hamiltonian dynamics \cite{D}. This provides many important probes to check the consistency of a theory, as listed below,
\begin{enumerate}
\item
The number of propagating degrees of freedom may be calculated in the phase space from the relation
\begin{eqnarray}
N = N_1 - 2N_2 - N_3 \label{dof}
\end{eqnarray}

where $N_1 =$ Total number of canonical variables, $N_2 =$ Total number of first class constraints 
and, 
 $N_3 = $Total number of second class constraints. Since the Chern Simons fields have no independent dynamics, the no. of degrees of freedom should be $N = 4 $ for our model. Physically, this corresponds to $\psi$ and $\psi*$ and their conjugate momenta.
 
 \item
 The number of primary first class constraints is equal to the number of independent gauge degrees of freedom. Note that this number can alternatively be obtained from the number of independent local symmetries of the action.
\end{enumerate}
Consistency in the Hamiltonian analysis is essential for a feasible model. We will see that the 
model (\ref{sl}) for the Schrodinger field coupled with non relativistic space is consistent from this point of view. This is remarkable because a host of models have been proposed for this problem, many of which have some differences with (\ref{sl}). Also it may be pointed out that Hamiltonian treatment of these theories are not much available.  

 In the following section we will discuss the Dirac approach to the constraint analysis of the problem.
\section{Canonical Analysis - the constraints of the theory}

To proceed with the canonical analysis of (\ref{wl}) we define the momenta $\pi$, $\pi^{*}$,$\pi^{0}_{\mu}$, $\pi^{a}_{k}$, $\pi^{\mu}_{ab}$, $\pi^{l}_{b0}$, $\pi^{0}_{a0}$ conjugate to the fields $\psi$, $\psi^{*}$,$\Sigma^{\mu}_{0}$, $\Sigma_{k}{}^{a}$, $B^{ab}_{\mu}$, $B^{b0}_{l}$, $\pi^{0}_{a0}$ respectively. Then
\begin{eqnarray}
\pi=\frac{\partial\mathcal{L}}{\partial\dot{\psi}} =\frac{Mi}{2}\psi^{*} \hspace{.2cm}; \hspace{.2cm}\pi^{*}=\frac{\partial\mathcal{L}}{\partial\dot{\psi^{*}}} =-\frac{Mi}{2}\psi \notag\\
\pi^{0}_{\mu}=\frac{\partial\mathcal{L}}{\partial\dot{\Sigma^{\mu}_{0}}}=0 \hspace{.2cm}; \hspace{.2cm} 
\pi^{a}_{k}=\frac{\partial\mathcal{L}}{\partial\dot{\Sigma^{k}_{a}}}=0  \hspace{.2cm}; \hspace{.2cm} \notag\\
\pi^{\mu}_{ab}=\frac{\partial\mathcal{L}}{\partial\dot{B^{ab}_{\mu}}}=0 \hspace{.2cm}; \hspace{.2cm}
\pi^{l}_{b0}=\frac{\partial\mathcal{L}}{\partial\dot{B^{b0}_{l}}}=\epsilon^{kl}\epsilon_{ab}\Lambda^{a}_{k} \notag\\
\pi^{0}_{a0}=\frac{\partial\mathcal{L}}{\partial\dot{B^{a0}_{0}}}= 0
\label{m}
\end{eqnarray}
The Poisson brackets (PB) between the canonical pairs are usual:
\begin{eqnarray}
{\{\psi(x),\pi(y)\}}=\delta^{2}(x-y)\notag\\
{\{\psi^{*}(x),\pi^{*}(y)\}}=\delta^{2}(x-y)\notag\\
{\{\Sigma^{\mu}_{0}(x),\pi^{0}_{\nu}(y)\}}=\delta^{\mu}_{\nu}\delta^{2}(x-y)\notag\\
{\{\Sigma^{l}_{b}(x),\pi^{a}_{k}(y)\}}=\delta^{a}_{b}\delta^{l}_{k}\delta^{2}(x-y)\notag\\
{\{B^{ab}_{\nu}(x),\pi^{\mu}_{cd}(y)\}}=\delta^{\mu}_{\nu}(\delta^{a}_{c}\delta^{b}_{d}-\delta^{b}_{c}\delta^{a}_{d})\delta^{2}(x-y)\notag\\
{\{B^{a0}_{k}(x),\pi^{l}_{b0}(y)\}}=\delta^{l}_{k}\delta^{a}_{b}\delta^{2}(x-y)\notag\\
{\{B^{b0}_{0}(x),\pi^{0}_{a0}(y)\}}=\delta^{b}_{a}\delta^{2}(x-y)
 \label{cpb}
\end{eqnarray}

From definition (\ref{m}) the following primary constraints emerge,
\begin{align}
\Omega_{1}=\pi-\frac{Mi}{2}\psi^{*}\thickapprox 0\hspace{.2cm};\hspace{.2cm}
\Omega_{2}&=\pi^{*}+\frac{Mi}{2}\psi \thickapprox 0\notag\\
\Omega^{0}_{\mu}=\pi^{0}_{\mu}\thickapprox 0\hspace{.2cm};\hspace{.2cm}
\Omega^{a}_{k}&=\pi^{a}_{k}\thickapprox 0\notag\\
\Omega^{\mu}_{ab}=\pi^{\mu}_{ab}\thickapprox 0\hspace{.2cm};\hspace{.2cm}
\Omega^{0}_{a0}&=\pi^{0}_{a0}\thickapprox 0\notag\\
\Omega^{l}_{b0}&=\pi^{l}_{b0}-\epsilon^{kl}\Lambda^{a}_{k}\epsilon_{ab} \thickapprox 0 \label{pc}
\end{align}

As is well known, conserving the primary constraints (\ref{cpb}) we may get secondary constraints.
We have to construct the total Hamiltonian, which is the canonical Hamiltonian improved by the linear combinations of the primary constraints.
The canonical Hamiltonian density of the theory is given by
\begin{eqnarray}
\mathcal{H}_{can}=\pi\dot{\psi}+\pi^{*}\dot{\psi^{*}}+\pi^{0}_{\mu}\dot{\Sigma^{\mu}_{0}}+\pi^{a}_{k}\dot{\Sigma^{k}_{a}}+\pi^{\mu}_{ab}\dot{B^{ab}_{\mu}}+\pi^{l}_{b0}\dot{B^{b0}_{l}}+\pi^{0}_{a0}\dot{B^{a0}_{0}}-\mathcal{L}
\label{canh}
\end{eqnarray}
Explicitly,
\begin{multline}
\mathcal{H}_{can}=-\frac{M}{\Sigma^{0}_{0}}\Bigl[\frac{i}{2}\Sigma^{k}_{0}\left(\psi^{*}\partial_{k}\psi-\psi\partial_{k}\psi^{*}\right)-\Sigma^{\mu}_{0}B^{a0}_{\mu}mx_{a}\psi^{*}\psi\\
-\frac{1}{2m}\Sigma^{k}_{a}\Sigma^{l}_{a}\left(\partial_{k}\psi^{*}\partial_{l}\psi+iB^{b0}_{l}mx_{b}\psi\partial_{k}\psi^{*}-iB^{b0}_{k}mx_{b}\psi^{*}\partial_{l}\psi+B^{c0}_{k}B^{b0}_{l}m^{2}x_{c}x_{b}\psi^{*}\psi\right)\Bigr]\\
+\epsilon^{kl}\Lambda^{0}_{0}\frac{\epsilon_{ab}}{2}\left(\partial_{k}B^{ab}_{l}-\partial_{l}B^{ab}_{k}+\frac{1}{2}B^{a0}_{k}B^{b0}_{l}\right)\\
-\epsilon^{kl}\Lambda^{a}_{0}\Bigl[\frac{\epsilon_{ab}}{2}\left(\partial_{l}B^{b0}_{k}-\partial_{k}B^{b0}_{l}\right)-\frac{\epsilon_{cd}}{4}\left(B^{a0}_{k}B^{cd}_{l}-B^{a0}_{l}B^{cd}_{k}\right)\Bigr]\\
+2\epsilon^{kl}\Lambda^{a}_{k}\Bigl[\frac{\epsilon_{ab}}{2}\partial_{l}B^{b0}_{0}-\frac{\epsilon_{cd}}{4}\left(B^{a0}_{0}B^{cd}_{l}-B^{a0}_{l}B^{cd}_{0}\right)\Bigr]
\end{multline}

The total Hamiltonian is

\begin{multline}
H_{T}=\int{d^{2}x}\left(\mathcal{H}_{can}+\lambda_{1}\Omega_{1}+\lambda_{2}\Omega_{2}+\lambda^{\mu}_{0}\Omega^{0}_{\mu}+\lambda^{k}_{a}\Omega^{a}_{k}+\frac{1•}{2•}\lambda^{ab}_{\mu}\Omega^{\mu}_{ab}+\lambda^{b0}_{l}\Omega^{l}_{b0}+\lambda^{a0}_{0}\Omega^{0}_{a0}\right)
\end{multline}

Here $\lambda_{1}$, $\lambda_{2}$, $\lambda^{\mu}_{0}$, $\lambda^{k}_{a}$, $\lambda^{ab}_{\mu}$, $\lambda^{b0}_{l}$, $\lambda^{a0}_{0}$ are Lagrange multipliers enforcing the constraints. In this theory, the non-vanishing fundamental Poisson brackets are given by

\begin{align*}
\label{pc1}
{\{\Omega_{1}(x),\Omega_{2}(y)\}}&=-iM\delta^{2}\left(x-y\right)\\
{\{\Omega_{1}(x),\Omega^{a}_{k}(y)\}}&=\frac{i\psi^{*}}{2}M\Lambda^{a}_{k}\delta^{2}\left(x-y\right)\\
{\{\Omega_{2}(x),\Omega^{a}_{k}(y)\}}&=-\frac{i\psi}{2}M\Lambda^{a}_{k}\delta^{2}\left(x-y\right)\\
{\{\Omega^{a}_{k}(x),\Omega^{l}_{b0}(y)\}}&=-\epsilon^{jl}\epsilon_{db}\Lambda^{a}_{j}\Lambda^{d}_{k}\delta^{2}\left(x-y\right)
\end{align*}
where we have used (\ref{cpb}). The primary constraints are denoted by the generic symbol $\Omega$, The index structure is sufficient to identify the particular one. Apparently, all the constraints
have nonzero PBs between each other, However, it may so happen that by combinations of the
constraints, a subset of them can be made to have vanishing PBs with all the elements of the set of constraints. For the time being let us carry on with 
the stationary of the primary constraints $\Omega^{0}_{a0}$ i.e; $\dot{\Omega}^{0}_{a0}={\{\Omega^{0}_{a0}(x),H_{T}\}} \thickapprox 0$ which yields the 
following expression,
\begin{equation}
\label{s15}
\Gamma_{a}=-Mmx_{a}\psi^{*}\psi+\epsilon^{kl}\epsilon_{da}\partial_{l}\left(\Lambda^{d}_{k}\right)+\frac{\epsilon^{kl}}{2}\Lambda^{a}_{k}\epsilon_{cd}B^{cd}_{l} \thickapprox 0 
\end{equation}
Note that the terms containing $x^a$ and the rest are separately zero. Two new secondary constraints are
thus obtained,
\begin{eqnarray}
\label{s}
\Phi_1 = \psi^{*}\psi\thickapprox 0 
\end{eqnarray}
and
\begin{equation}
\label{s11}
\Phi_{a}=\epsilon^{kl}\epsilon_{da}\partial_{l}\left(\Lambda^{d}_{k}\right)+\frac{\epsilon^{kl}}{2}\Lambda^{a}_{k}\epsilon_{cd}B^{cd}_{l} \thickapprox 0 
\end{equation}

The stationary of the primary constraint $\Omega^{0}_{ab}$ i.e; $\dot{\Omega}^{0}_{ab}={\{\Omega^{0}_{ab}(x),H_{T}\}} \thickapprox 0$ gives the secondary constraints as

\begin{equation}
\Phi_2 =\epsilon^{kl}\Lambda^{a}_{k}B^{a0}_{l} \thickapprox 0
\end{equation}

Conserving $\pi^{j}_{ef}$ in time, a secondary constraint emerges

\begin{equation}
S_{j}=\epsilon^{kj}\partial_{k}(\Lambda^{0}_{0})-\epsilon^{kj}\Lambda^{a}_{0}B^{a0}_{k}+\epsilon^{kj}\Lambda^{a}_{k}B^{a0}_{0} \thickapprox 0
\end{equation}

From $\dot{\pi}^{0}_{j}={\{\pi^{0}_{j}(x),H_{T}}\} \thickapprox 0$, we get further secondary constraints expression as,

\begin{equation}
\label{s12}
\Gamma_{j}^{'}=\frac{M_{i}}{2}\left(\psi^{*}\partial_{j}\psi-\psi\partial_{j}\psi^{*}\right)-MB^{a0}_{j}mx_{a}\psi^{*}\psi+\epsilon^{kl}\epsilon_{ab}\Lambda^{a}_{j}\partial_{k}B^{b0}_{l}+\frac{\epsilon^{kl}}{2}\epsilon_{cd}\Lambda^{a}_{j}B^{a0}_{k}B^{cd}_{l} \thickapprox 0
\end{equation}
Noting that the terms containing $x_a$ should be vanishing separately, we get a
 new secondary constraint, 
\begin{equation}
\bar{S_{k}}=\frac{Mi}{2}\left(\psi^{*}\partial_{k}\psi-\psi\partial_{k}\psi^{*}\right)-\epsilon^{jn}\epsilon_{da}B^{a0}_{k}\partial_{n}\Lambda^{d}_{j}+\epsilon^{jn}\epsilon_{ab}\Lambda^{a}_{k}\partial_{j}B^{b0}_{n} \thickapprox 0 
\end{equation}
where some simplification have been done using (\ref{s11},\ref{s12}).
Finally, conservation of $\pi_0^0 \thickapprox 0 $ leads to
\begin{multline}
\bar{\Gamma}=-\frac{M}{\Sigma^{0}_{0}}\Bigl[\frac{i}{2}\Sigma^{k}_{0}\left(\psi^{*}\partial_{k}\psi-\psi\partial_{k}\psi^{*}\right)-\Sigma^{k}_{0}B^{a0}_{k}mx_{a}\psi^{*}\psi\\
-\frac{1}{2m}\Sigma^{k}_{d}\Sigma^{l}_{d}{\{\partial_{k}\psi^{*}\partial_{l}\psi-iB^{a0}_{l}mx_{a}\left(\psi^{*}\partial_{k}\psi-\psi\partial_{k}\psi^{*}\right)+B^{a0}_{k}B^{b0}_{l}m^{2}x_{a}x_{b}\psi^{*}\psi}\}\\
+\epsilon^{kl}\epsilon_{ab}\Lambda^{0}_{0}\left(\partial_{k}B^{ab}_{l}+\frac{1}{4}B^{a0}_{k}B^{b0}_{l}\right)
+\Lambda^{a}_{0}\epsilon^{kl}\left(\epsilon_{ab}\partial_{k}B^{b0}_{l}+\frac{\epsilon_{cd}}{2}B^{a0}_{k}B^{cd}_{l}\right) \thickapprox 0\label{compo}
\end{multline}

Looking at (\ref{compo}) we see that it holds irrespective of $x^a$. But it can only happen if 
\begin{multline}
\bar{\Gamma}=-\frac{M}{\Sigma^{0}_{0}}\Bigl[\frac{i}{2}\Sigma^{k}_{0}\left(\psi^{*}\partial_{k}\psi-\psi\partial_{k}\psi^{*}\right)\psi\\
-\frac{1}{2m}\Sigma^{k}_{a}\Sigma^{l}_{a}{\{\partial_{k}\psi^{*}\partial_{l}\psi}\\
+\epsilon^{kl}\epsilon_{ab}\Lambda^{0}_{0}\left(\partial_{k}B^{ab}_{l}+\frac{1}{4}B^{a0}_{k}B^{b0}_{l}\right)
+\Lambda^{a}_{0}\epsilon^{kl}\left(\epsilon_{ab}\partial_{k}B^{b0}_{l}+\frac{\epsilon_{cd}}{2}B^{a0}_{k}B^{cd}_{l}\right) \thickapprox 0\label{compo1}
\end{multline}

and
\begin{multline}
\bar{\Gamma}=-\Sigma^{k}_{0}B^{a0}_{k}\psi^{*}\psi\\
-\frac{1}{2m}\Sigma^{k}_{a}\Sigma^{l}_{a}{\{\partial_{k}\psi^{*}\partial_{l}\psi-iB^{a0}_{l}\left(\psi^{*}\partial_{k}\psi-\psi\partial_{k}\psi^{*}\right)}\}\label{compo2}
\end{multline}

\ref{compo}) is equivalent to   (\ref{compo1})  and  (\ref{compo2}). Simplifying , we get two new set of constraints,
\begin{align}
S&=\frac{M}{2m}\Sigma^{k}_{c}\Sigma^{l}_{c}\partial_{k}\psi^{*}\partial_{l}\psi+\epsilon^{jn}\epsilon_{ab}\left(\partial_{j}B^{ab}_{n}+\frac{1}{4}B^{a0}_{j}B^{b0}_{n}\right) \thickapprox 0 \notag \\
S^{'}_{e}&=\Sigma^{k}_{c}\Sigma^{l}_{c}\epsilon^{jn}\epsilon_{fd}B^{e0}_{l}\left(B^{d0}_{k}\partial_{n}\Lambda^{f}_{j}-2\Lambda^{f}_{k}\partial_{j}B^{d0}_{n}-\frac{1}{2}\Lambda^{a}_{k}B^{a0}_{j}B^{fd}_{n}\right) \thickapprox 0 
\end{align}
Conserving the rest of the primary constraints $\Omega_1$, $\Omega_2$, $\Omega^a_k$, $\Omega^l_{b0}$ and the
new secondary constraints $\Gamma_a$, $\Gamma$, $\Gamma_{j}$, $\Gamma_{j}^{'}$, $\bar{\Gamma}$   no new constraints generate ; only some of the multipliers are fixed. The constraint structure is thus closed.

The secondary constraints are then listed below:

\begin{align}
\Phi_{1}&=\psi^{*}\psi \thickapprox 0 \notag \\
\Phi_{d}&=\epsilon^{kl}\epsilon_{ad}\partial_{l}\Lambda^{a}_{k}+\frac{\epsilon^{kl}}{2}\Lambda^{d}_{k}\epsilon_{ca}B^{ca}_{l} \thickapprox 0 \notag \\
\Phi_{2}&=\epsilon^{kl}\Lambda^{a}_{k}B^{a0}_{l} \thickapprox 0 \notag \\
S_{j}&=\epsilon^{kj}\partial_{k}\Lambda^{0}_{0}-\epsilon^{kj}\Lambda^{a}_{0}B^{a0}_{k}+\epsilon^{kj}\Lambda^{a}_{k}B^{a0}_{0} \thickapprox 0 \notag \\
\bar{S_{k}}&=\frac{Mi}{2}\left(\psi^{*}\partial_{k}\psi-\psi\partial_{k}\psi^{*}\right)-\epsilon^{jn}\epsilon_{da}B^{a0}_{k}\partial_{n}\Lambda^{d}_{j}+\epsilon^{jn}\epsilon_{ab}\Lambda^{a}_{k}\partial_{j}B^{b0}_{n} \thickapprox 0 \notag \\
S&=\frac{M}{2m}\Sigma^{k}_{c}\Sigma^{l}_{c}\partial_{k}\psi^{*}\partial_{l}\psi+\epsilon^{jn}\epsilon_{ab}\left(\partial_{j}B^{ab}_{n}+\frac{1}{4}B^{a0}_{j}B^{b0}_{n}\right) \thickapprox 0 \notag \\
S^{'}_{e}&=\Sigma^{k}_{c}\Sigma^{l}_{c}\epsilon^{jn}\epsilon_{fd}B^{e0}_{l}\left(B^{d0}_{k}\partial_{n}\Lambda^{f}_{j}-2\Lambda^{f}_{k}\partial_{j}B^{d0}_{n}-\frac{1}{2}\Lambda^{a}_{k}B^{a0}_{j}B^{fd}_{n}\right) \thickapprox 0 \label{sc}
\end{align}
The complete set of constraints of the theory comprises of (\ref{pc}) and  (\ref{sc}). The analysis of the constraints in first and second class gives a host of informations, as we have seen. We will now take up the issue.

\subsection{Classification  of the  constraints   and degrees of freedom count} In the Dirac method the constraints are divided in first and second class according to whether they have all mutual Poisson brackets vanishing or not. Using the fundamental Poisson brackets (\ref{cpb}) we can straightforwardly work out these brackets.
The non-vanishing Poisson brackets are given by-
\begin{align}
\label{split}
{\{{\Omega}_{1}(x),\Omega_2(y)\}}&=-iM\delta^{2}(x-y)\\
{\{\Omega_{1}(x),\Phi_{1}(y)}\}&=-\psi^{*}\delta^{2}(x-y)\\
{\{\Omega_{2}(x),\Phi_{1}(y)}\}&=-\psi\delta^{2}(x-y)\\
{\{\Omega_{1}(x),\bar{S_{k}}(y)}\}&=\frac{Mi}{2}\left[\partial^{y}_{k}\psi^{*}(y)\delta^{2}(x-y)-\psi^{*}(y)\partial^{y}_{k}\left(\delta^{2}(x-y)\right)\right]\\
{\{\Omega_{2}(x),\bar{S_{k}}(y)}\}&=\frac{Mi}{2}\left[\psi(y)\partial^{y}_{k}\left(\delta^{2}(x-y)\right)-\partial^{y}_{k}\psi(y)\delta^{2}(x-y)\right]\\
{\{\Omega_{1}(x),S(y)}\}&=-\frac{M}{2m}\Sigma^{k}_{c}\Sigma^{l}_{c}\partial^{y}_{k}\psi^{*}(y)\partial^{y}_{l}\left(\delta^{2}(x-y)\right)\\
{\{\Omega_{2}(x),S(y)}\}&=-\frac{M}{2m}\Sigma^{k}_{c}\Sigma^{l}_{c}\partial^{y}_{l}\psi(y)\partial^{y}_{k}\left(\delta^{2}(x-y)\right)\\
{\{\Omega^{a}_{k}(x),\Omega_{1}(y)}\}&=-\frac{i\psi^{*}}{2}M\Lambda^{a}_{k}\delta^{2}(x-y)\\
{\{\Omega^{a}_{k}(x),\Omega_{2}(y)}\}&=\frac{i\psi}{2}M\Lambda^{a}_{k}\delta^{2}
(x-y)
\end{align}
\begin{align}
\label{split1}
{\{\Omega^{0}_{0}(x),S_{j}(y)\}}&=\epsilon^{kj}\partial^{y}_{k}(\Lambda^{0}_{0}\Lambda^{0}_{0}\delta^{2}(x-y))-\epsilon^{kj}B^{a0}_{k}\Lambda^{0}_{0}\Lambda^{a}_{0}\delta^{2}(x-y)\\
{\{\Omega^{0}_{k}(x),S_{j}(y)\}}&=-\epsilon^{pj}B^{a0}_{p}\Lambda^{0}_{0}\Lambda^{a}_{k}\delta^{2}(x-y)\\
{\{\Omega^{a}_{k}(x),\Omega^{l}_{b0}(y)}\}&=-\epsilon^{pl}\epsilon_{cb}\Lambda^{c}_{k}\Lambda^{a}_{p}\delta^{2}(x-y) \\
{\{\Omega^{a}_{k}(x),\Phi_{d}(y)}\}&=\epsilon^{jl}\epsilon_{cd}\partial^{y}_{l}\left(\Lambda^{c}_{k}\Lambda^{a}_{j}\delta^{2}(x-y)\right)+\frac{1}{2}\epsilon^{jl}\epsilon_{cb}B^{cb}_{l}\Lambda^{d}_{k}\Lambda^{a}_{j}\delta^{2}(x-y) \\
{\{\Omega^{a}_{k}(x),\Phi_{2}(y)}\}&=\epsilon^{pl}B^{b0}_{l}\Lambda^{b}_{k}\Lambda^{a}_{p}\delta^{2}(x-y)\\
{\{\Omega^{a}_{k}(x),S_{j}(y)}\}&=\left[-\epsilon^{lj}B^{b0}_{l}\Lambda^{b}_{k}\Lambda^{a}_{0}+\epsilon^{lj}B^{b0}_{0}\Lambda^{b}_{k}\Lambda^{a}_{l}\right]\delta^{2}(x-y)
\end{align}
\begin{align}
\label{split2}
{\{\Omega^{a}_{k}(x),\bar{S}_{l}(y)}\}&=\frac{i}{2}\left(\psi^{*}\partial_{l}\psi-\psi\partial_{l}\psi^{*}\right)M\Lambda^{a}_{k}\delta^{2}(x-y)\\
&-\epsilon^{jn}\epsilon_{db}B^{b0}_{l}\partial^{y}_{n}\left(\Lambda^{d}_{k}\Lambda^{a}_{j}\delta^{2}(x-y)\right)+\epsilon^{jn}\epsilon_{cb}\partial_{j}B^{b0}_{n}\Lambda^{c}_{k}\Lambda^{a}_{l}\delta^{2}(x-y)\\
{\{\Omega^{a}_{k}(x),S(y)}\}&=\frac{M}{2m}\left[\Sigma^{j}_{c}\Sigma^{l}_{c}\Lambda^{a}_{k}\partial^{y}_{j}\psi^{*}\partial^{y}_{l}\psi-\Sigma^{j}_{a}\partial^{y}_{j}\psi^{*}\partial^{y}_{k}\psi-\Sigma^{l}_{a}\partial^{y}_{k}\psi^{*}\partial^{y}_{l}\psi\right]\delta^{2}(x-y)\\
{\{\Omega^{a}_{k}(x),\Phi_{e}(y)}\}&=-\epsilon^{jn}\epsilon_{fd}\Bigl[B^{e0}_{k}\Sigma^{p}_{a}\bigl(B^{d0}_{p}\partial^{y}_{n}\Lambda^{f}_{j}-2\Lambda^{f}_{p}\partial^{y}_{j}B^{d0}_{n}-\frac{1}{2}\Lambda^{b}_{p}B^{b0}_{j}B^{fd}_{n}\bigr)\\
&+B^{e0}_{l}\Sigma^{l}_{a}\left(B^{d0}_{k}\partial^{y}_{n}\Lambda^{f}_{j}-2\Lambda^{f}_{k}\partial^{y}_{j}B^{d0}_{n}-\frac{1}{2}\Lambda^{b}_{k}B^{b0}_{j}B^{fd}_{n}\right)\Bigr]\delta^{2}(x-y)\\
&+\epsilon^{jn}\epsilon_{fd}B^{e0}_{l}\Sigma^{p}_{c}\Sigma^{l}_{c}\Bigl[B^{d0}_{p}\partial^{y}_{n}\bigl(\Lambda^{f}_{k}\Lambda^{a}_{j}\delta^{2}(x-y)\bigr)\\
&-2\partial^{y}_{j}B^{d0}_{n}\Lambda^{f}_{k}\Lambda^{a}_{p}\delta(x-y)-\frac{1}{2}B^{b0}_{j}B^{fd}_{n}\Lambda^{b}_{k}\Lambda^{a}_{p}\delta^{2}(x-y)\Bigr]\\
{\{\Omega^{l}_{ab}(x),\Phi_{d}(y)}\}&=-\epsilon^{kl}\epsilon_{ab}\Lambda^{d}_{k}\delta^{2}(x-y)
\end{align}
\begin{align}
{\{\Omega^{l}_{ab}(x),S(y)}\}&=-2\epsilon^{jl}\epsilon_{ab}\partial^{y}_{l}\bigl(\delta^{2}(x-y)\bigr)\\
{\{\Omega^{l}_{ab}(x),S^{'}_{e}(y)}\}&=\Sigma^{k}_{c}\Sigma^{n}_{c}\epsilon^{jl}\epsilon_{ab}B^{e0}_{n}\Lambda^{d}_{k} 
B^{d0}_{j}\delta^{2}(x-y)\\
{\{\Omega^{l}_{b0}(x),\Phi_{2}(y)}\}&=-\epsilon^{kl}\Lambda^{b}_{k}\delta^{2}(x-y)\\
{\{\Omega^{l}_{b0}(x),S_{j}(y)}\}&=\epsilon^{lj}\Lambda^{b}_{0}\delta^{2}(x-y)\\
{\{\Omega^{l}_{b0}(x),\bar{S_{k}}(y)}\}&=\epsilon^{jn}\epsilon_{db}\partial^{y}_{n}\left(\Lambda^{d}_{j}\right)\delta^{l}_{k}\delta^{2}(x-y)-\epsilon^{jl}\epsilon_{ab}\Lambda^{a}_{k}\partial^{y}_{j}\left(\delta^{2}(x-y)\right)\\
{\{\Omega^{l}_{b0}(x),S(y)}\}&=-\frac{1}{2}\epsilon^{jl}\epsilon_{ab}B^{a0}_{j}\delta^{2}(x-y)\\
{\{\Omega^{0}_{a0}(x),S_{j}(y)}\}&=-\epsilon^{kj}\Lambda^{a}_{k}\delta^{2}(x-y)\\
\end{align}\\
\begin{multline*}
{\{\Omega^{l}_{bo}(x),S'_{e}(y)}\}=-\Sigma^{l}_{c}\Sigma^{p}_{c}\epsilon^{jn}\epsilon_{fb}B^{e0}_{p}\partial^{y}_{n}\Lambda^{f}_{j}\delta^{2}(x-y)\\
+2\Sigma^{k}_{c}\Sigma^{p}_{c}\epsilon^{jl}\epsilon_{fb}B^{e0}_{p}\Lambda^{f}_{k}\partial^{y}_{j}\left(\delta^{2}(x-y)\right)\\
+\frac{•1}{•2}\Sigma^{k}_{c}\Sigma^{p}_{c}\epsilon^{ln}\epsilon_{fd}B^{e0}_{p}\Lambda^{b}_{k}B^{fd}_{n}\delta^{2}(x-y)\\
-\Sigma^{k}_{c}\Sigma^{l}_{c}\epsilon^{jn}\epsilon_{fd}\delta^{e}_{b}\left(B^{d0}_{k}\partial^{y}_{n}\Lambda^{f}_{j}-2\Lambda^{f}_{k}\partial^{y}_{j}B^{d0}_{n}-\frac{1}{2•}\Lambda^{a}_{k}B^{a0}_{j}B^{fd}_{n}\right)\delta^{2}(x-y)\\
\end{multline*}

Poisson bracket of $\Omega^{0}_{l} \approx 0$ vanishes with all the constraints except $S_{j}$.
\begin{equation}
{\{\Omega^{0}_{l}(x),S
_{j}(y)}\}=-\epsilon^{kj}B^{a0}_{k}\Lambda^{0}_{0}\Lambda^{a}_{l}\delta^{2}(x-y)
\end{equation}

If 
we construct,
\begin{eqnarray}
\bar{\Omega^{0}_{l}}=\pi^{0}_{l}-\Lambda^{0}_{0}B^{a0}_{l}\pi^{0}_{a0} \thickapprox 0
\end{eqnarray}, 
then
\begin{eqnarray}
{\{\bar{\Omega^{0}_{l}}(x),S_{j}(y)}\}=\epsilon^{kj}\Lambda^{0}_{0}\left(\Lambda^{a}_{k}B^{a0}_{l}-\Lambda^{a}_{l}B^{a0}_{k}\right)\delta^{2}(x-y) \approx 0
\end{eqnarray}
where we have used $\Lambda^{a}_{l}B^{a0}_{k}=\Lambda^{a}_{k}B^{a0}_{l}$ which is obtained from constraint $\Phi_{2}$.
Also $\bar{\Omega^{0}_{l}}$ has vanishing Poisson brackets with all other constraints. Replacing ${\Omega^{0}_{l}}$ by $\bar{\Omega^{0}_{l}}$ in the set of constraints (\ref{pc}.\ref{sc}) we find that
 $\bar{\Omega_k{}^0}$
, $\Omega_{ab}{}^0$ have vanishing PBs among themselves and with other constraints. With these results the classification of the constraints can easily be done.
The complete classification of constraints is summarized in Table \ref{tab:Classification of Constraints} below.
\begin{table}[ht]
\caption{Classification of Constraints}
\label{tab:Classification of Constraints}
\centering
\begin{tabular}{c c c}
\hline\hline\\
\ & First Class & Second Class \\
 \hline\\
 Primary & $\bar{\Omega_k{}^0}$,   ${\Omega^{0}_{ab}}$ & ${\Omega_{1}}$,${\Omega_{2}}$, ${\Omega^{0}_{0}}$, $\Omega^{a}_{k}$, $\Omega^{l}_{ab}$, $\Omega^{l}_{b0}$, $\Omega^{0}_{a0}$ \\[1em]
\hline\\
 Secondary &  &  $\Phi_{1}$, $\Phi_{d}$, $\Phi_{2}$, $S_{j}$, $\bar{S_{k}}$, $S$, $S^{'}_{e}$ \\[1em]
 \hline\hline
\end{tabular}
\end{table}

The results tabulated above can be physically interpreted in the following way:
\begin{enumerate}
\item

 The number of independent fields is 18. That gives 36 fields in the phase space as each field is accompanied with its canonically conjugate momentum. The number of first class constraints is 3 while the number of secondary constraints is 26. The number of independent degrees of freedom in the phase space can now be calculated.
 Using \ref{dof} we get $N = 36 - 2 \times 3 - 26 =4  $. So, the no. of degrees of freedom in configuration space is 2. Physically, they correspond to $\psi$ and $\psi^*$. Note that the Chern Simons dynamics does not contribute any propagating degree of freedom.
 
 \item 
 
 The number of independent primary first class constraints is three. According to Dirac conjecture
 it is the number of independent 'gauge' degrees of freedom. Here arbitrary functions in the solutions of the equations of motion will then be three in number. Physically, these are the consequence of three local symmetry operations, one rotation and two boosts.
\end{enumerate}
\section{Canonical analysis with $\Sigma_0{}^k$ = 0}
We have already discussed at few places in this paper that the motivation of our work is to check the consistency of the model (\ref{wl}) and to posit it in relation to the corresponding actions obtained from other approaches. To our knowledge the latter  are of the  same form as that of \cite{SW}. This form differs from our model in essence by the absence of the term $ \Sigma_0{}^k$ = 0. It will then be crucial to check whether in our model we substitute $\Sigma_0{}^k$ = 0 it still has the same
physically consistent Hamiltonian structure.

 We therefore consider the truncated model
\begin{multline}
\label{wlt}
\mathcal{L} = M{}\Bigl[\frac{i}{2}\left(\psi^{*}\partial_{0}\psi-\psi\partial_{0}\psi^{*}\right))\\
-B^{a0}_{0}mx_{a}\psi^{*}\psi-\frac{1}{2m}\Sigma^{k}_{a}\Sigma^{l}_{a}\left(\partial_{k}\psi^{*}-iB^{a0}_{k}mx_{a}\psi^{*}\right)\left(\partial_{l}\psi+iB^{a0}_{l}mx_{a}\psi\right)\Bigr]\\
-\epsilon^{kl}\frac{\epsilon_{ab}}{2}\left(\partial_{k}B^{ab}_{l}-\partial_{l}B^{ab}_{k}+\frac{1}{2}B^{a0}_{k}B^{b0}_{l}\right)
-2\epsilon^{kl}\Lambda^{a}_{k}\Bigl[\frac{\epsilon_{ab}}{2}\left(\partial_{l}B^{b0}_{0}-\partial_{0}B^{b0}_{l}\right)-\frac{\epsilon_{cd}}{4}\left(B^{a0}_{0}B^{cd}_{l}-B^{a0}_{l}B^{cd}_{0}\right)\Bigr]
\end{multline}
which is obtained from (\ref{wl}) by putting $ \Sigma_0{}^k$ = 0 in it. We have also taken  $ \Sigma_0{}^0$ = 1 as it is possible when there is no transformation of time i.e. there is spatial diffeomorphism only \cite{BMM1}.The canonical analysis proceeds in the same way as above.

Performing the canonical analysis, we obtain the following primary constraints:
\begin{align*} 
\Omega_{1}&=\pi-\frac{Mi}{2}\psi^{*} \approx 0\\
\Omega_{2}&=\pi^{*}+\frac{Mi}{2}\psi \approx 0\\
\Omega^{a}_{k}&=\pi^{a}_{k} \approx 0\\
\Omega^{\mu}_{ab}&=\pi^{\mu}_{ab} \approx 0\\
\Omega^{0}_{a0}&=\pi^{0}_{a0} \approx 0\\
\Omega^{l}_{b0}&=\pi^{l}_{b0}-\epsilon^{kl}\epsilon_{ab}\Lambda^{a}_{k} \approx 0\\
\end{align*}
The stationary of the primary constraints $\Omega^{\mu}_{ab}$ and $\Omega^{0}_{a0}$ give the following secondary constraints:
\begin{align*}
\Phi_{1}&=\psi^{*}\psi \approx 0\\
\Phi_{d}&=\epsilon^{kl}\epsilon_{ad}\partial_{l}(\Lambda^{a}_{k})+\frac{\epsilon^{kl}}{2}\Lambda^{d}_{k}\epsilon_{ca}B^{ca}_{l} \approx 0\\
\Phi_{2}&=\epsilon^{kl}\Lambda^{a}_{k}B^{a0}_{l} \approx 0\\
S^{'}_{j}&=\epsilon^{kj}\Lambda^{a}_{k}B^{a0}_{0} \approx 0\\
\end{align*}
The iteration terminates with the closure of the constraint algebra.

The non-vanishing poisson brackets between the constraints are given by
\begin{align*}
{\{\Omega_{1}(x),\Omega_{2}(y)}\}&=-Mi\delta^{2}(x-y)\\
{\{\Omega_{1}(x),\Phi_{1}(y)}\}&=-\psi^{*}\delta^{2}(x-y)\\
{\{\Omega_{2}(x),\Phi_{1}(y)}\}&=-\psi\delta^{2}(x-y)\\
{\{\Omega^{a}_{k}(x),\Omega^{l}_{b0}(y)}\}&=-\epsilon^{pl}\epsilon_{cb}\Lambda^{c}_{k}\Lambda^{a}_{p}\delta^{2}(x-y)\\
{\{\Omega^{a}_{k}(x),\Omega_{1}(y)}\}&=-\frac{i\psi^{*}}{2}M\Lambda^{a}_{k}\delta^{2}(x-y)\\
{\{\Omega^{a}_{k}(x),\Omega_{2}(y)}\}&=\frac{i\psi}{2}M\Lambda^{a}_{k}\delta^{2}(x-y)\\
{\{\Omega^{a}_{k}(x),\Phi_{d}(y)}\}&=\epsilon^{jl}\epsilon_{cd}\partial^{y}_{l}\left(\Lambda^{c}_{k}\Lambda^{a}_{j}\delta^{2}(x-y)\right)\\
&+\frac{1}{2}\epsilon^{jl}\epsilon_{cb}B^{cb}_{l}\Lambda^{d}_{k}\Lambda^{a}_{j}\delta^{2}(x-y)\\
{\{\Omega^{a}_{k}(x),\Phi_{2}(y)}\}&=\epsilon^{pl}B^{b0}_{l}\Lambda^{b}_{k}\Lambda^{a}_{p}\delta^{2}(x-y)\\
{\{\Omega^{l}_{b0}(x),\Phi_{2}(y)}\}&=-\epsilon^{kl}\Lambda^{b}_{k}\delta^{2}(x-y)\\
{\{\Omega^{0}_{a0}(x),S^{'}_{j}(y)}\}&=-\epsilon^{kj}\Lambda^{a}_{k}\delta^{2}(x-y)\\
{\{\Omega^{a}_{k}(x),S^{'}_{j}(y)}\}&=\epsilon^{pj}\Lambda^{a}_{p}\Lambda^{b}_{k}B^{b0}_{0}\delta^{2}(x-y)\\
{\{\Omega^{l}_{ab}(x),\Phi_{d}(y)}\}&=-\epsilon^{kl}\epsilon_{ab}\Lambda^{d}_{k}\delta^{2}(x-y)\\
\end{align*}
The complete classification of constraints is summarized in Table 2 below.
\begin{table}[ht]
\caption{Classification of Constraints when $\Sigma_0{}^k = 0$}
\label{tab:Classification of Constraints of the truncated model}
\centering
\begin{tabular}{c c c}
\hline\hline\\

\ & First Class & Second Class \\
 \hline\\
 Primary &  ${\Omega^{0}_{ab}}$ & ${\Omega_{1}}$, ${\Omega_{2}}$, $\Omega^{a}_{k}$, $\Omega^{l}_{ab}$, $\Omega^{l}_{b0}$, $\Omega^{0}_{a0}$ \\[1em]
\hline\\
 Secondary &  & $\Phi_{1}$, $\Phi_{d}$, $\Phi_{2}$, $S^{'}_{j}$ \\[1em]
 \hline\hline
\end{tabular}
\end{table}
The number of fields is 15, the number of first class constraints is one whereas there are 20 secondary constraints. So the number of degrees of freedom in the phase space is 8. This is twice as large as the physical degrees of freedom. So we see that the model with $\Sigma_0{}^k = 0$ is unable to give the hamiltonian analysis consistently.

Again, we see from table -2 that the number of primary first class constraints is one. So the model predicts one local symmetry as opposed to three physical symmetries. So taking $\Sigma_0{}^k =0$ also gives incorrect symmetries. Further
investigation shows that the boost symmetries are lost. This connection with boost is indeed remarkable, not only for GGT but also in general.

 In the above we have assumed $\Sigma_0{}^0 = 1$ in addition to $\Sigma_0{}^k = 0$. One may enquire the reason behind such choice. It has been proved in \cite{BMM1} that for spatial diffeomorphism where time translation parameter is zero, 
 $\Sigma_0{}^0 $ is a constant which can conveniently put to be unity. The condition $\Sigma_0{}^k = 0$ {\bf{is not permitted}} by GGT in general. So there is no wonder
 that it leads to unphysical result. The assertion can be verified by taking  .$\Sigma_0{}^0 $ in account following similar calculation we can show that the number of degrees of freedom comes out to be three, different from the physical value. 
 Moreover, from the  point of view of lost symmetries, there is no improvement.
\section{Discussion of the results} The basic issue discussed in this paper is the consistency of  a non relativistic complex scalar field (the Schrodinger field) coupled with background gravity by Galilean gauge theory (GGT)\cite{BMM1}, \cite{BM4}  in phase space.The dynamics of gravity is assumed to be given by the Chern - Simons gravity action. The  model is invariant under spatial diffeomorphism. The pioneering model in this field was given in \cite{SW}. However, it was riddled with certain difficulties concerning symmetries. The solution provided in \cite{SW} was to exploit certain relationship between the gauge and boost parameters. The same model was derived in \cite{J} from a relativistic theory in the $c\to \infty$ limit. But that raised several questions like the reason for the reduction of independent number of symmetry parameters (owing to the equality of gauge and boost parameter) and more important, what would happen if one likes to couple a free Schrodinger field with background gravity \cite{n1}? The confusions were correctly diagonosed to be due to the lack of understanding the proper way to couple with the nonrelativistic Newton cartan spacetime. Thus it was proposed that the gauge field be included in the elements of NC algebra \cite{J}. However, to many it appears little contrived. Certainly, the masters who erected the structure of NC spacetime never conjectured it. Also this proposal is not free of inner problems (like the issue of connection etc.).
GGT was developed in this background \cite{BMM1}, \cite{BMM2} which followed an alternative approach based on localisation of symmetry. Equation (\ref{sl}) is our
result for a non relativistic complex scalar field (the Schrodinger field) coupled with background gravity in Newton - Cartan spacetime.


 In GGT it is pretty straightforward to specialize (\ref{sl}) so that it is invariant under spatial diffeomorphism and include a gauge field in the action.,
 From (\ref{sl})
 \begin{multline}
\label{sl}
\mathcal{L}={\sqrt{g}}\Bigl[\frac{i}{2}\left(\psi^{*}\partial_{0}\psi-\psi\partial_{0}\psi^{*}\right)+\frac{i}{2}\Sigma^{k}_{0}\left(\psi^{*}\partial_{k}\psi-\psi\partial_{k}\psi^{*}\right)\\
-B_{0}\psi^{*}\psi-\Sigma^{k}_{0}B_{k}\psi^{*}\psi-\frac{1}{2m}\Sigma^{k}_{a}\Sigma^{l}_{a}\left(\partial_{k}\psi^{*}-iB_{k}\psi^{*}\right)\left(\partial_{l}\psi+iB_{l}\psi\right)\Bigr]
\end{multline}
where we have substituted $\Sigma_0{}^0 =1$. The spatial metric is defined as
\begin{equation}
g_{ij} = \Lambda_i{}^a\Lambda_j^a
\end{equation} 
Clearly $M=\det {\Lambda_i{}^a} = \sqrt{g}$ where $g= \det{g_{ij}}$

 Now the gauge field can be simply included by replacing the partial derivatives by the appropriate covariant derivative 
\begin{eqnarray}
{D}_{0}\phi&=\partial_0\phi +iA_0\phi\notag\\
{D}_{k}\phi&=\partial_k\phi +iA_k\phi \label{P110}
\end{eqnarray}
where $A_\mu$ is an (external) gauge field
The resulting model can be organnised as \cite{BMM3, BM4}.
\begin{eqnarray}
\tilde{S} &=& \int dx^0 d^2x \sqrt{g}[ \frac{i}{2}\left(\phi^{*}{\bar{D}}_{0}\phi-\phi{\bar{D}}_0\phi^{*}\right) -g^{kl}\frac{1}{2m}\bar{D}_k\phi^{*}\bar{D}_l\phi]
\nonumber\\&+&\int dx^0 d^2x \sqrt{g} [\frac{i}{2}\Sigma_0{}^k\left(\phi^{*}{\bar{D}}_{k}\phi
-\phi{\bar{D}}_k\phi^{*}\right)]
\label{diffaction12}
\end{eqnarray}
where
\begin{eqnarray}
\bar{D}_{0}\phi&=\partial_0\phi +i\bar{{A_0}\phi}\notag\\
\bar{D}_{k}\phi&=\partial_k\phi +i\bar{A_k}\phi \label{P1100}
\end{eqnarray}
and
\begin{eqnarray}
\bar{A_\mu} = A_\mu + B_\mu 
\end{eqnarray}

Compare
(\ref{P1100}) with 
 the action  given by\cite{SW}
\begin{equation}\label{free-L1}
  S = \int dx^0 dx \sqrt{g}\left[\frac{i}{2} (\phi^{*}D_{0}\phi-\phi D_{0}\phi^*)
    - \frac{g^{ij}}{2m}(D_i\phi^*D_j\phi)\right],
\end{equation}

The differences between (\ref{free-L1}) and  (\ref{diffaction12}) is in the former the spin connections $B_\mu{}^{ab}$ and $B_\mu{}^{a0}$ are absent. Since the Schrodinger field is a $3$- scalar $B_\mu{}^{ab}$ is dropped but the same is not true for  $B_\mu{}^{a0}$. However, the principal difference is the absence of the  of the term containing $\Sigma_0{}^k$ in the action.
The Hamiltonian analysis in the first place confirms the models in the phase space 
To check the impact of the  we  have repeated the Hamiltonian analysis of our model, this time taking $\Sigma_0{}^k = 0$. 
 We have seen that by dropping $\Sigma_0{}^k$= 0, we no longer
get a consistent theory. Hence the model (\ref{free-L1}) is ruled out due to its inconsistency in the phase space. We conclude that the model given by GGT must be taken as \ref{free-L1}. As for the model of $\cite{SW}$, we note that a Hamiltonian analysis of it is unavailable.

\section{Conclusion}
A nonrelativistic diffeomorphism invariant Schrodinger field theory coupled with Chern Simons gravity \cite{witten} has been considered. The 'matter '
 part of the theory has been obtained using the algorithm of the recently proposed Galilean gauge theory \cite{BMM1, BMM2, BMM3, BM4} which leads to coupling with gravity through the vierbeins and spin connections of the spacetime manifold. The gravity dynamics is given by the CS term which is an interesting alternative to (being equivalent to ) the Einstein Hilbert action  in $2+1$ dimensions \cite{blago}. The 
Schrodinger field theory coupled with background gravity was recently found to be very useful in connection with the research in fractional quantum Hall effect \cite{SW}. The model of \cite{SW} were used in diverse problems \cite{SW,Bekaert:2011qd,Hoyos:2011ez, Schaefer:2013oba,Hoyos:2013qna} but there were many loose ends ofit Thus, the metric transformed in an anomalous way and the Galilean symmetry could only be retrieved in the flat limit by equating the gauge and boost parameters.
The Chern Simons term which was known to be instrumental in FQHE was found to be incompatble with the NRDI of the model \cite{Hoyos:2011ez}. These problems were eradicated in the systematic treatment of GGT where the Schrodinger field theory coupled with background NC gravity was systematically obtained which  have \cite{BMM1, BMM2,BMM3, BM4}.
\begin{enumerate}
\item non relativistic spatial diffeomorphism invariance;
\item galilean symmetry in the flat limit
\item facility to include Chern Simons term as easily as any gauge interaction

\end{enumerate}

. As the Schrodinger field coupled with NC gravity is associated with very important phenomenologies, the details of it is required to be investigated from different points of view. The results of the present Hamiltonian analysis has demonstrated that not only the GGT model is physically consistent, any deviation from it would lead to unphysical conclusions.

We have performed a Hamiltonian analysis of spatially diffeomorphic nonrelativistic Schrodinger field theory coupled with Chern Simons gravity . The coupled model was derived from the recently developed Galilean gauge theory \cite{BMM1, BMM2,BMM3, BM4}. We have shown that the number of degrees of freedom matches with the physically expected values. Also, the number of independent gauge symmetries comes out to be same as the number of independent symmetries of the action. The coupled action contains a term which vanishes if the time space part of the vielbein in  Galilean coordinates is taken to be zero. We have explicitly worked out the constraint algebra of the reduced form but it failed to give correct values of the degrees of freedom and the independent symmetries of the truncated action.
Our results confirm that the model obtained from GGT is consistent in phase space
in its entirety, notwithstanding the difference with the other approaches. Also such analysis is not quite available in the literature. Also, it introduces a model with Chern - Simons gravity in the literature in this field

{\bf{acknowledgement}}
One of the authors (PM)
likes to thank Rabin Banerjee for useful discussions.

\end{document}